%% file: main.tex
\documentclass[conference]{IEEEtran}
\usepackage[utf8]{inputenc}
\usepackage[T1]{fontenc}

\usepackage{booktabs} % For formal tables

\usepackage[caption=false,font=footnotesize]{subfig}
\usepackage{fancyhdr}
\usepackage{comment}
\usepackage{mdwlist}
\usepackage{graphicx}
\usepackage{url}
\usepackage{color}
\usepackage{multirow}
\usepackage{amsmath}        % enables "cases" for piece-wise functions

\graphicspath{{incl/}}

\makeatletter
\def\blfootnote{\xdef\@thefnmark{}\@footnotetext}
\makeatother

\usepackage{setspace}
\def\smallerspacecaption{\vspace{-2mm}}
\def\smallspaceenum{\vspace{1.25mm}}

\IEEEoverridecommandlockouts

\begin{document}

\title{\huge 3D Integration: Another Dimension Toward Hardware Security}

\author{Johann Knechtel, Satwik Patnaik, and Ozgur Sinanoglu\\
	{\small Tandon School of Engineering, New York University, New York, USA}\\
	{\small Division of Engineering, New York University Abu Dhabi, United Arab Emirates}\\
	{\small \{johann, sp4012, ozgursin\}@nyu.edu}
\thanks{This work was supported in part by NYUAD under REF Grant RE218 and by NYU/NYUAD Center for Cybersecurity (CCS).}
}

\maketitle

\renewcommand{\headrulewidth}{0.0pt}
\thispagestyle{fancy}
\pagestyle{fancy}
\cfoot{
	\vspace{-1cm}
\copyright~2019 IEEE.
This is the author's version of the work. It is posted here for your personal use.
	Not for redistribution.\\
	The definitive Version of Record is published in
	Proc. IEEE Int.\ On-Line Test Symp.\ (IOLTS), 2019.
}

\begin{abstract}
We review threats and selected schemes concerning hardware security at design and manufacturing time as well as at runtime.  We find that 3D
integration can serve well to enhance the resilience of different hardware security schemes, but it also requires thoughtful use of the options
provided by the umbrella term of 3D integration.  Toward enforcing security at runtime, we envision secure 2.5D system-level integration of
untrusted chips and ``all around'' shielding for 3D ICs.
\end{abstract}
\smallspaceenum

\section{Introduction}
\label{sec:introduction}

Given the fact that integrated circuits (ICs) are at the heart of ubiquitous information technology, IC designers and vendors should seek to
establish trust into their products by all available means. However, doing so is a practical challenge as related efforts deviate from
the typically security-unaware design and manufacturing flows.
For example, it has been shown that the speculative execution in processors, which is an
industry-wide best practice, can be exploited to leak sensitive data~\cite{kocher18}.
Besides such concerns regarding data at runtime, the field of hardware security also spans other design- and manufacturing-time threats
such as reverse engineering (RE), intellectual property (IP) piracy, overproduction, or insertion of hardware Trojans (HTs)~\cite{BRS17}.
The latter threats arise due to outsourcing, which is a predominant trend for IC supply chains,
    as it is the case for many other industries nowadays.

Aside from traditional 2D IC manufacturing, research and development for 3D integration has made significant progress over recent years.
3D integration means to stack and interconnect multiple chips or active layers. There are two main drivers for 3D
integration~\cite{knechtel16_Challenges_ISPD,knechtel17_TSLDM}: (1)~the CMOS
scalability bottleneck, which becomes more exacerbated for advanced nodes by issues like routability, pitch scaling, and process variations;
and (2)~the need to advance means for heterogeneous and system-level integration.
Both drivers are also known as (1)~``More Moore'' and (2)~``More than Moore.''
Various studies, prototypes, and commercial products have shown that 3D integration can indeed offer significant benefits over conventional 2D
ICs, e.g., see \cite{fick13, iyer15,
	kim12_3dmaps, shilov18}.

\begin{figure}[tb]
\centering
\includegraphics[width=\columnwidth]{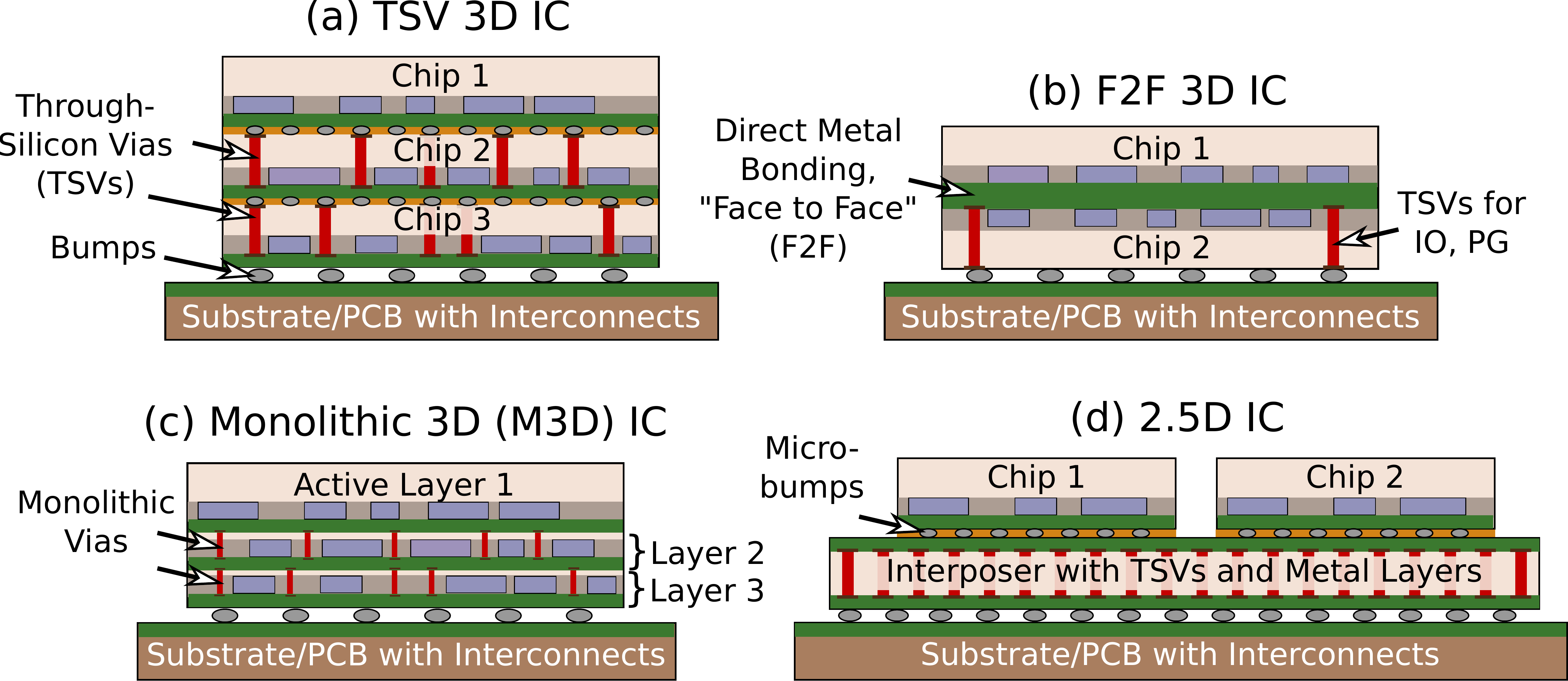}
\smallerspacecaption
\smallerspacecaption
\smallerspacecaption
\caption{Four flavors of 3D integration. Metal layers are colored in green, active layers in brown, along with logic in blue,
	and bonding layers in yellow.
\label{fig:3D_schemes}
}
\smallerspacecaption
\end{figure}

The umbrella term of 3D integration comprises four different flavors as follows
(Fig.~\ref{fig:3D_schemes}):
\begin{itemize}
\item[(a)] Through-silicon via (TSV) 3D ICs, where multiple chips are fabricated separately and then stacked and bonded. The vertical
interconnects across the 3D ICs are realized by relatively large metal TSVs which are cutting through the individual chips.
\item[(b)] Face-to-face (F2F) 3D ICs, where two chips are fabricated separately and then bonded directly at their back-end-of-line (BEOL) metal
``faces.''
TSVs or wirebonds are commonly used for external connections.
\item[(c)] Monolithic 3D (M3D) ICs, where multiple active layers are manufactured sequentially. The vertical interconnects are implemented by
regular metal vias.
\item[(d)] 2.5D ICs, where chips are fabricated separately and then stacked and bonded onto a system-level interconnect carrier, the so-called
interposer. This interposer can be either passive, comprising only metal layers and possibly some discrete devices, or active, containing some
logic.
\end{itemize}

In this paper, we discuss prior art concerning hardware security in general.  We then elaborate how 3D integration offers unique
opportunities to advance such schemes, along with a review of recent studies.

\section{Protection of Design IP}

Independent of 2D IC manufacturing or modern 3D integration, various techniques have been proposed over the years toward protection of chip
design IP.
Most works fall under one of the following three categories: logic locking (LL), layout camouflaging (LC), or split manufacturing (SM).  All
three schemes consider different threats; LL is concerned about untrusted end-users and malicious foundries, SM about untrusted foundries, and LC
about untrusted end-users.
The interested reader may also want to see~\cite{knechtel19_IP_COINS} for more details.

Now, 3D integration can serve to advance these schemes in different ways.  The main benefit provided by 3D integration is the \emph{physical
	separation} of components, as dictated by the security-enforcing designer and/or vendor, be it across interconnects, active devices, or
both. In Table~\ref{tab:prior_3D_design_man}, we provide an overview on selected works, which are also discussed next.

\begin{table}[tb]
\centering
\scriptsize
\caption{Selected 3D Schemes Targeting at Manufacturing Time}
\label{tab:prior_3D_design_man}
\smallerspacecaption
\setlength{\tabcolsep}{1.5mm}
\begin{tabular}{|c|c|c|c|c|c|}
\hline
\textbf{Reference} & 
\textbf{Style} & 
\textbf{Scheme} & 
\textbf{Scope} & 
\textbf{Assets} &
\textbf{Trusted Facility} \\ 
\hline 
\hline
~\cite{xie17} & 2.5D & SM & IP Piracy & Wires & BEOL \\ \hline
~\cite{imeson13} & 2.5D & SM & HT Prevent. & Wires & BEOL \\ \hline
~\cite{gu18} & F2F & SM \& LC & IP Piracy & Gates \& Wires & FEOL \& BEOL \\ \hline
~\cite{yan17_camo} & M3D & LC & IP Piracy & Gates & FEOL \\ \hline
~\cite{patnaik18_3D_ICCAD} & F2F & SM \& LC & IP Piracy & Gates \& Wires & BEOL \\ \hline
\end{tabular}
\smallerspacecaption
\smallerspacecaption
\end{table}

\subsection{Logic Locking}
To the best of our knowledge, 3D integration has not been explored yet for LL.  In a loosely related work by
Sengupta \emph{et al.}~\cite{sengupta19_LL_SM_DATE}, the authors leverage formal principles pertaining to LL in order to advance the notion
of SM. More specifically, they lock the FEOL and delegate the unlocking to a separate, trusted BEOL facility, namely by implementing the
LL key via BEOL-level routing toward fixed-logic drivers.
The authors note that their scheme can also be realized at the package or board level, which suggests an
implementation as 2.5D IC.

\subsection{Camouflaging}
Yan \emph{et al.}~\cite{yan17_camo} were first to propose LC dedicated for 3D integration, more specifically for monolithic 3D ICs.  The authors
developed and characterized custom libraries, and they evaluated their scheme
at both the cell and the
chip scale.
The device-level camouflaging is realized by dummy contacts, which has been proposed already previously for LC in classical 2D ICs.
Thus, while conceptionally not new, the work in~\cite{yan17_camo} leverages the benefits provided by monolithic 3D ICs, in an effort
to advance the scalability of LC. That is noteworthy because prior art for 2D-centric LC may incur excessive PPA cost. For example, the
2D NAND-NOR-XOR primitive of~\cite{rajendran13_camouflage} would incur 5.5$\times$ power, 1.6$\times$ delay, and 4$\times$ area cost
compared to a regular NAND gate.\footnote{The excessive cost of 2D LC schemes would arguably allow only for few gates being camouflaged.
This, in turn, renders prior art either fully prone to analytical attacks, e.g.,
see the Boolean satisfiability (SAT) framework in~\cite{yu17}, or it calls for advanced, SAT-resilient schemes. By nature, however, such schemes are low-corruptibility ones, thereby enabling
an attacker to obtain at least an approximate version of the 
IP~\cite{shamsi17}.}
In contrast, Yan \emph{et al.}~\cite{yan17_camo} report on average 25\% power cost, 15\% delay cost, and 43\% area \emph{savings} compared to regular
2D gates.

\subsection{Split Manufacturing}
As indicated by the terminology, SM means to split the manufacturing flow,
typically into an untrusted FEOL process and a trusted BEOL process thereafter.
For the FEOL facility, a split layout appears like a ``sea of gates,'' making it
difficult to infer the complete netlist readily. Still, since physical-design tools work holistically on both FEOL and BEOL, various traces can
well remain in the FEOL
\cite{wang16_sm,li19_SM_ML_DAC}.

To advance SM, leveraging 3D integration is straightforward and also promising.  That is because 3D integration allows to split
a design into multiple chips, which can maintain their FEOL and BEOL layers independently, whereas the 2.5D/3D stack can comprise further parts
of the system-level interconnects.
This system-level nature of 3D SM allows to manufacture, test, and withhold various functional components from untrusted parties, all as need be.
Moreover, concerns regarding the practicability of classical 2D SM can be elevated due to this very fact
that the 2.5D/3D stack can hold additional system-level interconnects independently of the regular FEOL and BEOL processing of individual chips.

The idea of 3D SM was outlined already in 2008, by Tezzaron Semiconductor Corp.~\cite{tezzaron08}.
Various early studies were hinting at 3D SM as well, but most had some limitations.  For example, Dofe \emph{et al.}~\cite{dofe17}
remained on the conceptional level. Xie \emph{et al.}~\cite{xie17} and Imeson \emph{et al.}~\cite{imeson13} utilized 2.5D integration with
``only'' wires being hidden from untrusted facilities---in principle, this is equivalent to traditional SM for 2D ICs but, as indicated, it is 
more practical. Still, the studies~\cite{xie17,imeson13} suffer from considerable layout cost.

Later on, DeVale
\emph{et al.}~\cite{devale17}, Gu \emph{et al.}~\cite{gu18}, and Patnaik \emph{et al.}~\cite{patnaik18_3D_ICCAD} explored SM in the
context of 3D ICs, effectively promoting ``native 3D SM.''
One key findings of those later studies is that both the partitioning as well as the design of the vertical interconnects play an important role
and define a cost-security trade-off as follows. The more the design is split across multiple chips, the higher the layout cost can become (also
		due to the need for more vertical interconnects), but the more flexibility an security-enforcing designer has to separate
components and thereby ``dissolve'' the IP across the 3D stack.

\subsection{Split Manufacturing and Camouflaging in Conjunction}
Both Gu \emph{et al.}~\cite{gu18} and Patnaik \emph{et al.}~\cite{patnaik18_3D_ICCAD} proposed 3D SM in conjunction with LC.
While Gu \emph{et al.}~\cite{gu18}~consider regular, FEOL-centric LC, Patnaik \emph{et al.}~\cite{patnaik18_3D_ICCAD}~argue that another approach is more appropriate, namely
the obfuscation of the vertical interconnects.

\begin{figure}[tb]
\centering
\includegraphics[width=.98\columnwidth]{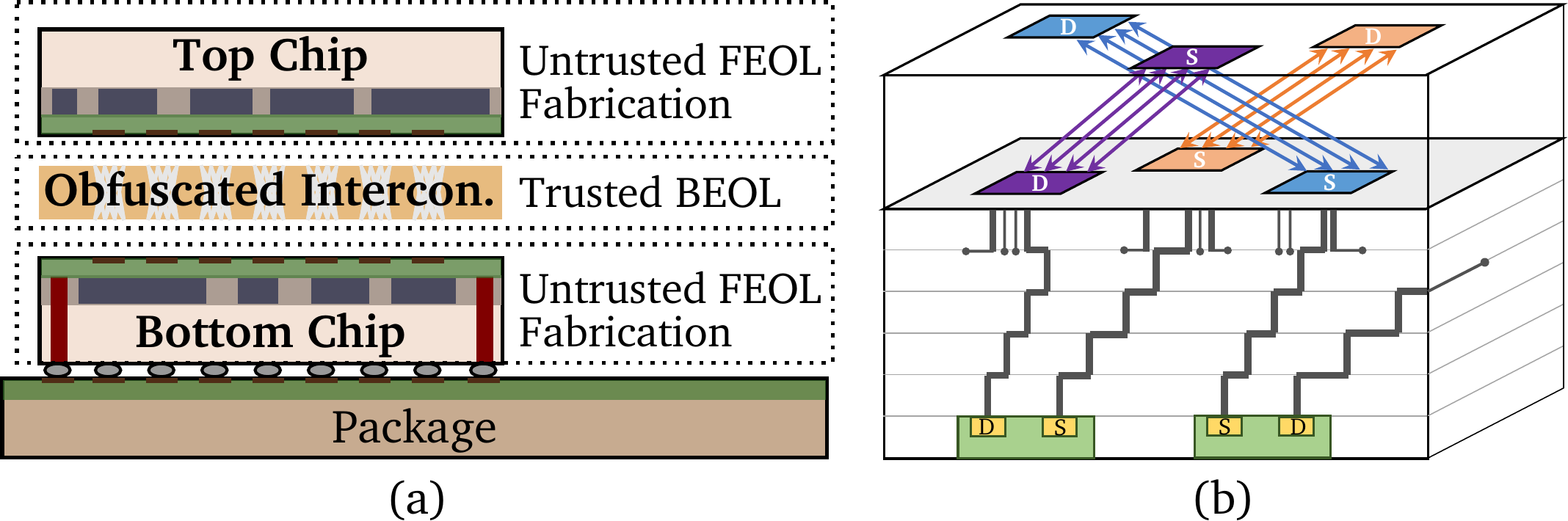}
\smallerspacecaption
\caption{Key concepts for 3D split manufacturing with camouflaging, as proposed in~\cite{patnaik18_3D_ICCAD}.
(a)~Obfuscated interconnects are added to render RE more difficult.
(b)~Randomized routing to hinder FEOL-foundry-based adversaries from inferring the netlist. For simplicity, only the bottom chip and the
additional, trusted BEOL interconnects are indicated.
\label{fig:3D_SM_LC}
}
\smallerspacecaption
\end{figure}

In their work~\cite{patnaik18_3D_ICCAD}, the authors propose to include additional metal
layers as redistribution layers (RDLs) between the chips of an F2F 3D IC.
As illustrated in Fig.~\ref{fig:3D_SM_LC}, these additional layers comprise
(a)~obfuscated interconnects (without loss of generality using magnesium- and magnesium-oxide-based vias) to render RE of the 3D IC more
difficult,
and (b)~randomized routing paths such that regular stacking of the chips will not readily reveal these missing interconnects.
By doing so, their work addresses both malicious foundries and end-users, along with affordable layout and
manufacturing cost~\cite{patnaik18_3D_ICCAD}.

Besides LC along with 3D SM as outlined above,
other works suggest camouflaging at the system level.
More specifically, Dofe \emph{et al.}~\cite{dofe16} propose to obfuscate the vertical communication links in 3D ICs
by rerouting within dedicated network-on-chip structures (NoCs) ``sandwiched'' between the chips.
In that sense, their idea is similar but more flexible to the randomized routing outlined above for Patnaik \emph{et
	al.}~\cite{patnaik18_3D_ICCAD}, although it is also more costly as it requires dedicated active layers instead of only metal layers.

\section{Hardware Security at Runtime}

Aside from the need for protecting the design IP and
rendering the globalized supply chain more trustworthy, there are
also various security risks arising at runtime. The major threats are (i)~unauthorized access or modification of data and (ii)~invasive
probing or modification of the hardware or its behavior.
For (i), this can occur via malicious software, HTs, side-channel attacks, misuse of test infrastructures, etc.
For (ii), this comprises scenarios like focused-ion beam milling, monitoring of photon emission, fault injection, etc.

Similar to the case of IP protection, 3D integration can also advance schemes focused on security at runtime, again by the virtue of
physical separation.
   Next, we provide an overview on selected works, and we also propose some novel concepts.

\subsection{Monitoring or Verification of Untrusted ICs}

In general, various monitoring or verification schemes have been proposed toward continuous control of ICs, e.g.,
	see~\cite{kim11_trojan,bhunia13,chandrasekharan15,wahby16}.  The common objective of these works is to detect any malicious or unexpected
		behaviour at runtime, emanating from software, hardware, or even both.

Extending such schemes via 3D integration is particularly promising. That is because the security-critical components
can be implemented separately using a trusted fabrication process and 3D-integrated later on with the commodity chip to be
monitored~\cite{wahby16,mysore06,valamehr13}.
Still, we caution that the physical implementation
can become a vulnerability by itself.
In~\cite{valamehr13}, for example, the authors propose ``introspective interfaces'' which, however, require additional logic within the commodity chip to
be monitored. It is easy to see that
these interfaces could fail when they are modified by any malicious actor involved with the design or manufacturing of that commodity
chip.  Thus,
a undesirable dependency arises, possibly thwarting the scheme altogether.
We note that the authors themselves acknowledge this limitation for their work in~\cite{valamehr13}.

\begin{figure}[tb]
\centering
\includegraphics[width=.728\columnwidth]{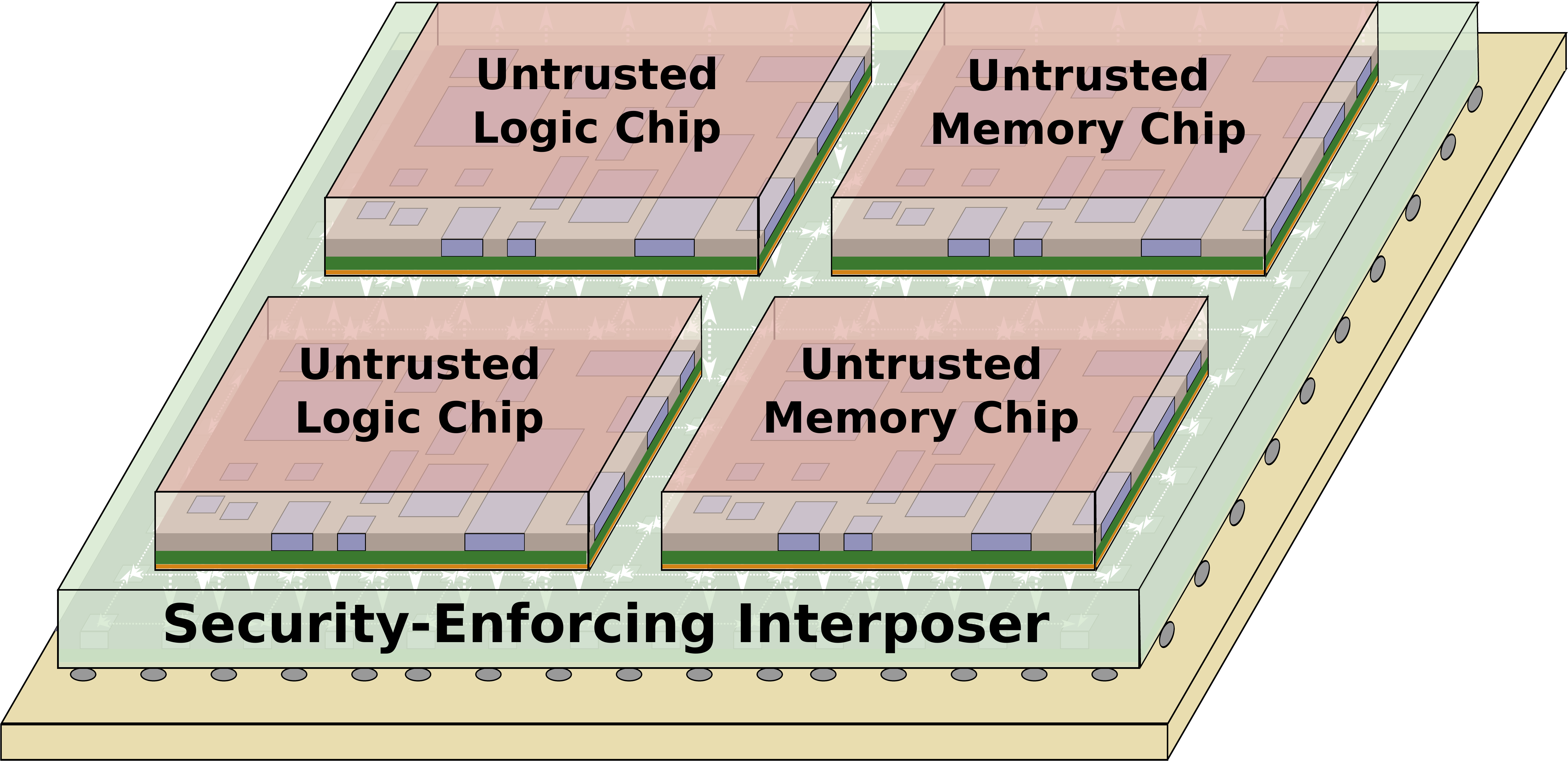}
\smallerspacecaption
\caption{Our concept for a secure 2.5D system-level integration of untrusted chips. All communication between chips and memories are monitored by
	the security features residing in the active, security-enforcing interposer.
\label{fig:3D_ROT}
}
\smallerspacecaption
\smallerspacecaption
\end{figure}

Here we envision an alternative for secure monitoring at runtime, based on 2.5D integration (Fig.~\ref{fig:3D_ROT}).
The essence of our approach is to provide a proper, system-level separation of untrusted commodity and trusted security components.
That is, all the physical interfaces and security features are delegated exclusively to an active interposer, which also serves as system-level
interconnect backbone and integration carrier.
Active interposer have been successful demonstrated, e.g., see~\cite{clermidy16}; they
can also be implemented using an old but trusted facility, thereby possibly allowing for cost savings.

\subsection{Probing, Monitoring, or Circuit Modification Attacks}

Semi- or fully-invasive probing, monitoring, fault injection, or even circuit modifications are arguably the most severe threats for hardware
security at runtime. Related attacks and various countermeasures have been demonstrated for classical 2D ICs, e.g.,
	 see~\cite{helfmeier13,tajik17_CCS,wang17_probing} and \cite{briais12,cioranesco14,yi16,weiner18}.

\begin{figure}[tb]
\centering
\includegraphics[width=.8\columnwidth]{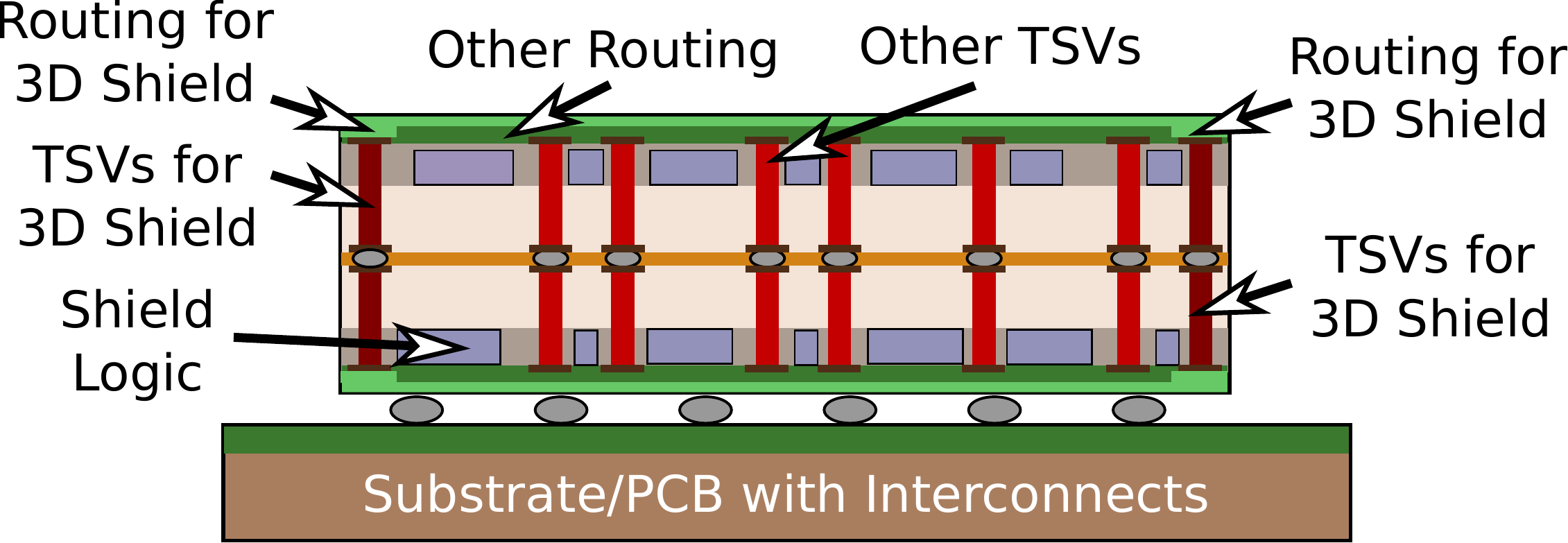}
\smallerspacecaption
\caption{Our proposal for ``all around'' shielding in 3D ICs.  The most suitable integration style is back-to-back (B2B) stacking,
	since it allows to protect the substrate backsides from invasive probing and non-invasive readout attacks.  The frontsides comprise
		dedicated layers/routing paths for the shield (light green), which are complemented by shield TSVs (dark red) placed at the IC
		boundaries.  The shield logic is integrated along with regular logic.
\label{fig:3D_cage}
}
\smallerspacecaption
\smallerspacecaption
\end{figure}

Despite the severity of those threats on the one hand, we also emphasize that, on the other hand, the benefits which 3D integration can offer to
protect against such threats are outstanding. To the best of our knowledge, 3D integration is the only enabler for an ``all around''
shield approach.
That is, only within 3D ICs we can build up a fully enveloping 3D shield structure
(Fig.~\ref{fig:3D_cage}).
Such a 3D shield would comprise (i)~dedicated metal layers/wires building up individual shields for both the chips,
and (ii)~TSVs to interconnect the shields of the two chips.
Toward the logical and physical design of the individual shields, prior art can be leveraged readily, e.g., see~\cite{cioranesco14,weiner18}.
Besides hindering probing attacks, we believe that such a 3D shield could also hinder other non-invasive but powerful attacks, e.g., monitoring
of the photon emission~\cite{tajik17_CCS}.

Similar protection has been discussed before in~\cite{briais12,cioranesco14}. However, the true potential for 3D shields has not been explored
	yet; the authors of those early studies considered techniques like B2B stacking
either as out-of-scope or as impractical, although without providing any substantiated critic.

\subsection{Side-Channel Attacks and Hardware Trojans}

Conducting a side-channel attack (SCA) means to carefully examine the physical emanations of an IC under attack, in order to extract some sensitive
information.  SCAs are powerful and hard to
prevent since any electronic device is inevitably subject to physical side-channels emissions at runtime.  For example, it has been shown that
the timing of caches or power consumption can be exploited to infer secret keys~\cite{zhou05}.

On the one hand, prior art considers SCAs which are targeted explicitly for 3D ICs.  For example, Gu \emph{et al.}~\cite{gu16_ICCD} and
Knechtel and Sinanoglu~\cite{knechtel17_TSC_DAC} seek to hinder thermal SCAs on 3D ICs at runtime and design time, respectively, whereas Dofe
\emph{et al.}~\cite{dofe17-CPA} seek to hinder power SCAs on 3D ICs.  On the other hand, some studies leverage the benefits provided by 3D
integration to apply security techniques otherwise considered too costly. For example, Bao and Srivastava~\cite{bao15} impose
random eviction and differing latencies across a cache architecture. The authors show that such techniques incur high performance cost in classical
2D ICs but can be realized even with some performance gains in a 3D IC.

Mossa \emph{et al.}~\cite{mossa17} have cautioned that HTs can become more stealthy and effectively tailored for 3D ICs than for 2D ICs.
The authors explore thermal triggers in detail, motivated by the fact that thermal management is a well-known challenge for 3D ICs by itself.
Finally, in a similar but general manner, we like to caution that the broader landscape of suppliers and actors involved with 3D integration can
open up new opportunities for attackers to embed different types of HTs.

\section{Summary}

In this short paper, we have reviewed major threats and selected schemes concerning hardware security at design/manufacturing time as well
as at runtime.  We note that 3D integration serves well to enhance different approaches for hardware security,
   but it also requires careful use of those novel 3D techniques.
   We have also outlined two advanced schemes for enforcing security at runtime, one based on 2.5D system-level integration of untrusted
   commodity chips, and one based on ``all around'' 3D shielding.

\input{main.bbl}

\end{document}